\documentclass[reprint,superscriptaddress,floatfix
 amsmath,amssymb,
 aps,
]{revtex4-1}
\usepackage{xcolor}
\usepackage{graphicx}% Include figure files
\usepackage{dcolumn}% Align table columns on decimal point
\usepackage{bm}% bold math
\usepackage{hyperref}% add hypertext capabilities
\usepackage[normalem]{ulem}
\usepackage{braket}
\usepackage{soul}

\newcommand{\ktn}{$^{39}$K}
\newcommand{\affINO}{\affiliation{CNR-INO, Istituto Nazionale di Ottica, 50019 Sesto Fiorentino, Italy}}
\newcommand{\affLENS}{\affiliation{European Laboratory for Nonlinear Spectroscopy - LENS, 50019 Sesto Fiorentino, Italy}}
\newcommand{\affUNIFI}{\affiliation{Dipartimento di Fisica e Astronomia, Universit\`{a} di Firenze, 50019 Sesto Fiorentino, Italy}}

\begin{document}

\preprint{APS/123-QED}

\title{Multimode Trapped Interferometer with Ideal Bose-Einstein Condensates}

\author{Leonardo Masi}
%\affiliation{Istituto Nazionale di Ottica, CNR-INO, 50019 Sesto Fiorentino, Italy}
\affINO
\affLENS

\author{Tommaso Petrucciani}
\affLENS
\affUNIFI

\author{Alessia Burchianti}
\affINO
\affLENS

\author{Chiara Fort}
\affINO
\affLENS
\affUNIFI

\author{Massimo Inguscio}
\affINO
\affLENS
\affiliation{Dipartimento di Ingegneria, Campus Bio-medico Universit\`{a} di Roma, 00128 Roma, Italy}

\author{Lorenzo Marconi}
\affINO

\author{Giovanni Modugno}
\affINO
\affLENS
\affUNIFI

\author{Niccolò Preti}
\affUNIFI

\author{Dimitrios Trypogeorgos}
\affiliation{CNR Nanotec, Institute of Nanotechnology, 73100 Lecce, Italy}

\author{Marco Fattori}
\affINO
\affLENS
\affUNIFI

\author{Francesco Minardi}
\affINO
\affLENS
\affiliation{Dipartimento di Fisica e Astronomia, Università di Bologna, 40127 Bologna, Italy}

\date{\today}% It is always \today, today,

\begin{abstract}
We experimentally demonstrate a multi-mode interferometer comprising a Bose-Einstein condensate of \ktn\ atoms trapped in a harmonic potential, where the interatomic interaction can be cancelled exploiting Feshbach resonances. Kapitza-Dirac diffraction from an optical lattice coherently splits the BEC in multiple momentum components equally spaced that form different interferometric paths, closed by the trapping harmonic potential. We investigate two different interferometric schemes, where the recombination pulse is applied after a full or half oscillation in the confining potential. We find that the relative amplitudes of the momentum components at the interferometer output are sensitive to external forces, through the induced displacement of the harmonic potential with respect to the optical lattice. We show how to calibrate the interferometer, fully characterize its output and discuss perspective improvements.
\end{abstract}

%\keywords{Suggested keywords}%Use showkeys class option if keyword
                              %display desired
\maketitle

%\tableofcontents
%\section{\label{sec:level1}First-level heading:\protect\\ The line
%break was forced \lowercase{via} \textbackslash\textbackslash}

Developed over three decades, atom interferometry represents the state-of-the-art for the measurements of accelerations \cite{peters1999measurement,Cronin_Interf_Review_2009} and rotations \cite{Gustavson_PrecisionRotation_1997} with unparalleled resolution, for the precise determination of fundamental constants \cite{clade2006determination,rosi2014precision, parker_measurement_2018} and for testing the equivalence principle \cite{gaaloul2010quantum,Asenbaum_2020}. In these interferometers, laser-cooled atomic samples are split in two separate paths and recombined by multiple photons transitions. Since most of those experiments are based on free-falling atomic samples the 
resolution typically scales as the square of the interrogation time, i.e. linearly with the length of the interferometer. However, on one hand, lengthy interferometers are technically demanding, since they require managing the spread of atomic wavepackets and  controlling external perturbations over large regions of space; on the other hand, their spatial resolution is obviously limited. 

Interferometry with trapped atoms offers the distinct advantage of extended interrogation times in  
compact setups, measuring forces and local fields with spatial resolution of a few micrometers. Several experiments have been performed using Bose-Einstein condensates (BEC) trapped in magnetic traps \cite{Tiecke_2003, pandey2019hypersonic}, generated near the surface of an atom chip \cite{schumm2005matter}, and in optical double-well potentials \cite{albiez2005direct, Spagnolli2017}. In these works the main effort was directed
toward engineering the external potential used to coherently split and recombine 
the wavefunction in two spatially separated modes.
In addition to exploiting long interrogation times, trapped atom interferometers can increase the interferometric phase in a direct way by enlarging the spatial separations between the paths.  
This goal has been pursued by different methods, e.g. by implementing double-well potentials with separations of several microns \cite{Spagnolli2017}, by coupling Wannier-Stark states that are several lattices sites distant \cite{Pelle2013}, by 
holding the two paths in a vertical lattice separated by large distances \cite{Charriere2012}, or by splitting a single condensate in the two traps formed by a transverse mode of an optical cavity \cite{Naik2018}. A recent proposal has also suggested to exploit the spread of the wavefunction during the Bloch dynamics in a horizontal lattice in the  presence of a weak force as a simple way to increase the spatial separations of the atoms and consequently increase the sensitivity of the interferometer \cite{Nalecz2020}. 

The present work demonstrates a new interferometric method for trapped quantum gases based on a multimode configuration, with more than two interferometric paths closed by a harmonic potential, where the coherent splitting and recombination of a BEC into multiple momentum components are realized by means of Kapitza-Dirac (KD) diffraction from a pulsed optical lattice. The method was proposed theoretically for non-interacting systems \cite{Li_Multimode_2014}.
A trapped KD interferometer has been 
implemented experimentally but only in the Mach-Zender two-mode configuration \cite{Wang2005, Sapiro_KD_2009}, with interacting $^{87}$Rb atoms for which 
the interpretation of the interference fringes is complicated by the interatomic interactions \cite{Burchianti_effect_2020}. Like in the above mentioned interferometers, also our sensitivity on the measured acceleration increases with the spatial separation of the paths; it reaches a level allowing 
the detection of the beam pointing instability of the optical trap.

More specifically, we report here on the realization of a horizontal multi-mode trapped interferometer, where 
a nearly ideal Bose-Einstein condensate %of \ktn\  atoms 
is KD diffracted by an optical lattice 
into 
components with momenta equal to multiples of the lattice wavevector $k$, i.e. with $p=m\, \hbar k$, with $m$ integer. A KD pulse initiates the oscillation of the different momentum orders which, after half an oscillation in the harmonic trap, %are recombined 
return to the initial position with opposite momenta and are recombined using another KD pulse.
Importantly,  
all the
momentum components spatially recombine at the trap minimum only if the potential is harmonic over their oscillation amplitude. This is easily the case for magnetic traps generated by macroscopic coils \cite{Sapiro_KD_2009}, but it represents a tight constraint for optical dipole traps (ODT's), that are the most common choice when, e.g., the control of interactions through Feshbach resonances is sought.
Here we use an optical trap and take advantage of a large-spacing ($\sim 5\,\mu$m) optical lattice, that reduces the recoil velocity, hence the oscillation amplitude, by a factor of 10, with respect to the commonplace lattice spacing of $0.5 \mu$m. Specifically, we create the periodic potential exploiting a recently developed technique, named ``beat-note superlattice'' \cite{Beatnote_2021}
capable of realizing lattices with a large effective period in a retro-reflected configuration, with laser wavelengths of the order of 1 $\mu$m.
In order to investigate experimentally the operation of the KD interferometer in the presence of external perturbations, we first observe the evolution of the different diffracted orders in the ODT, confirming a harmonic and symmetric evolution. Then, we calibrate the output distribution as a function of the time interval and the phase difference between the two pulses after a full oscillation. These measurements serve to set the right parameters during the operation. We then apply a controlled horizontal force through a magnetic field gradient and measure the momentum distribution after half an oscillation, for different values of the gradient field.
We observe a clear dependence of the atomic populations in the different momentum components on the external force and we compare the observed results with analytical predictions.

\section{Theoretical analysis}

A KD interferometer in a harmonic trap of frequency $\omega$ detects an unknown acceleration $a$ through the induced displacement of the trap minimum, $d=a/\omega^2$, measured with respect to the wavefront of the applied optical lattice. 
%The displacement is encoded in the phase factors imprinted by the KD pulses to each momentum component, equal to $e^{i m \phi}$, with $\phi= k d$, on the component of momentum $m \hbar k$ ($m$ integer). 
The displacement is encoded in the phase factor $e^{im\phi}$, with $\phi=kd$, imprinted by the KD pulses on the different momentum components $m \hbar k$.

The basic working of the interferometer can be best visualized considering the case with only three momentum components $m=0,\pm 1$ and focusing on $m=+1$. 
After the first KD pulse, the half oscillation in the trap reverses the momentum of all components. The momentum component $m=+1$ emerges from the second KD pulse through two distinct paths: (A) the $m=-1$ component is generated by the first KD pulse, its momentum is reversed into $m=+1$ by the half-period evolution and then it is left unperturbed by the second KD pulse; (B) the $m=0$ component emerging from the first KD pulse is scattered into $m=+1$ by the second. On path A the phase factor $e^{-i\phi}$ is imprinted by the first KD pulse, while on path B $e^{+i\phi}$ is imprinted by the second KD pulse: their sum produces interference on the momentum population at the interferometer output. 

Indeed, like in any other light-pulse interferometer \cite{Hogan_Light-pulse_2009} the interferometric phase accumulated on each path is obtained as the sum of the terms originated during the light-atom interaction and terms due to the evolution between the light pulses. The latter are given by the action along the classical trajectory, which vanishes whenever the time separation between the two light pulses equals an integer multiple of the harmonic half-period, independently from the initial momentum.

Taking into account all the momentum components, the wavefunction at the interferometer output is exactly calculated \cite{Li_Multimode_2014}. We consider that the initial wavefunction $\psi_0(x)$ corresponds to the ground state of the harmonic oscillator, that the lattice potential during the two KD pulses is given by $V_i(x) = V_0 \sin(k x + \phi_i)\, (i=1,2)$, and the pulse duration $\delta t$ is so short that only the wavefunction phase is affected (Raman-Nath limit \cite{Huckans_2009}). Then, the wavefunction at the interferometer output after half oscillation is:
\begin{eqnarray}
    \psi_{H}(x) &=& \psi_0(-x) e^{-i 2 \beta \sin\phi \cos(kx + \delta)}\nonumber \\
    &=& \psi_0(-x)  \sum_m J_m\left( 2\beta \sin\phi\right) (-i)^m e^{i m k x} e^{i m \delta}
    \label{eq:psi_half}
\end{eqnarray}
where $\beta=V_0 \delta t/\hbar$, $\phi=(\phi_1+\phi_2)/2$ and $\delta=(\phi_2-\phi_1)/2$ are the average and semi-difference of the the phases of the two pulses, and $J_m$ denotes the Bessel function of order $m$. Clearly a displacement $d$ of the harmonic potential is equivalent to a translation of both phases by $kd$. 

The acceleration, that gives rise to the displacement $d$, can be measured considering various observables constructed from the output populations. 
One notable example, on which we will focus in the following, is the fraction of atoms remaining in  the initial $m=0$ component, $O_1=N_0/N$: due to the multimode interference, varying $d$ this observable is expected to display a 
peak that narrows as the number of interfering components increases. The half-width-at-half-maximum (HWHM) phase, corresponding to $J_0^2(2\beta \sin\phi_{\mathrm{HWHM}})=1/2$, is $\phi_{\mathrm{HWHM}} = k d_{\mathrm{HWHM}} \simeq 0.56/\beta$. This is similar to light interference with multiple beams occurring, e.g., in a high-finesse optical cavity: in the very same way, the resolution of the KD interferometer increases with $\beta$ at the expense of its dynamic range \cite{Li_Multimode_2014}. More specifically,
the resolution of the measured acceleration $a$ %depends on the derivative of the chosen experimental observable. Using observable $O_1$, we have:
is
\begin{eqnarray}
    \delta a &=& \left|\frac{dO_1}{da}\right|^{-1} \Delta O_1 \nonumber\\ 
    &=& |2 J_0(2\beta\sin\phi) J_1(2\beta\sin\phi)  2\beta \cos\phi |^{-1} \frac{\omega^2}{k} \Delta O_1
   \label{eq:delta_a}
\end{eqnarray}
where $\Delta O_1$ is the experimental uncertainty associated with the observable. The resolution %sensitivity 
is maximum, i.e. $\delta a$ is minimum, when the lattice position is such that $\phi$ is close to an integer multiple of $\pi$. In this case \footnote{The function $|4 J_0(2\beta\sin\phi) J_1(2\beta\sin\phi) \cos\phi|$ takes a maximum value approximately equal to 1.2, weakly dependent on $\beta$.},
\begin{equation}
    \delta a \simeq \frac{\omega^2}{k} \frac1{\beta} \Delta O_1,
    \label{eq:min_delta_a}
\end{equation}
the resolution increases with $\beta$, that is proportional to the number of momentum components significantly populated, showing the benefit of multimode interference. 
This is consistent with the analysis based on the Fisher information and the Cramer-Rao bound \cite{Li_Multimode_2014, cheng_theory_2016}. Experimentally, the $O_1$ peak can be centered at any chosen value of acceleration by controlling the position of the minima of the lattice, extending the dynamic range of this kind of interferometer for the measurements of small forces around $a=0$.
%This is a notable difference with respect to other interferometric schemes that exploit cold atoms in vertical optical lattices \cite{poli_precision_2011} to precisely measure the gravity acceleration. % In the following, considering such experiments, we refer for sake of brevity, to their sensitivity alone% 

In alternative to $O_1$, we can fit the measured values of the fractional populations at the interferometer output with the squared Bessel functions, having the phase $\phi$ as the single fit parameter. This approach lacks a clear analogy with multiple-beams optical interferometers, but has the advantage of using all populations on the same footing. From a practical point of view we verified that the two approaches yield the same sensitivity.

Finally, for the following it is useful to derive also the output wavefunction for an interferometer where the two KD pulses are separated by a full-period evolution:
\begin{equation}
    \psi_{F}(x) %&=\psi_0(x) e^{-i 2 \beta \sin ( k x  + \phi) \cos \delta  }\nonumber \\
    = \psi_0(x) \sum_m J_m\left( 2 \beta \cos\delta \right) e^{-i m k x} e^{-i m \phi}, 
    \label{eq:psi_full}
\end{equation}
showing that the populations of the momentum components in this full-period interferometer are sensitive only to the {\it relative} displacement of the two KD pulses, i.e. to $\delta$ and not to $\phi$. We will exploit this property to calibrate the displacement of the lattice and investigate its stability. \\

\section{Experimental setup}

\begin{figure}[!htbp]
\begin{center}
    \includegraphics[width=1.0\columnwidth]{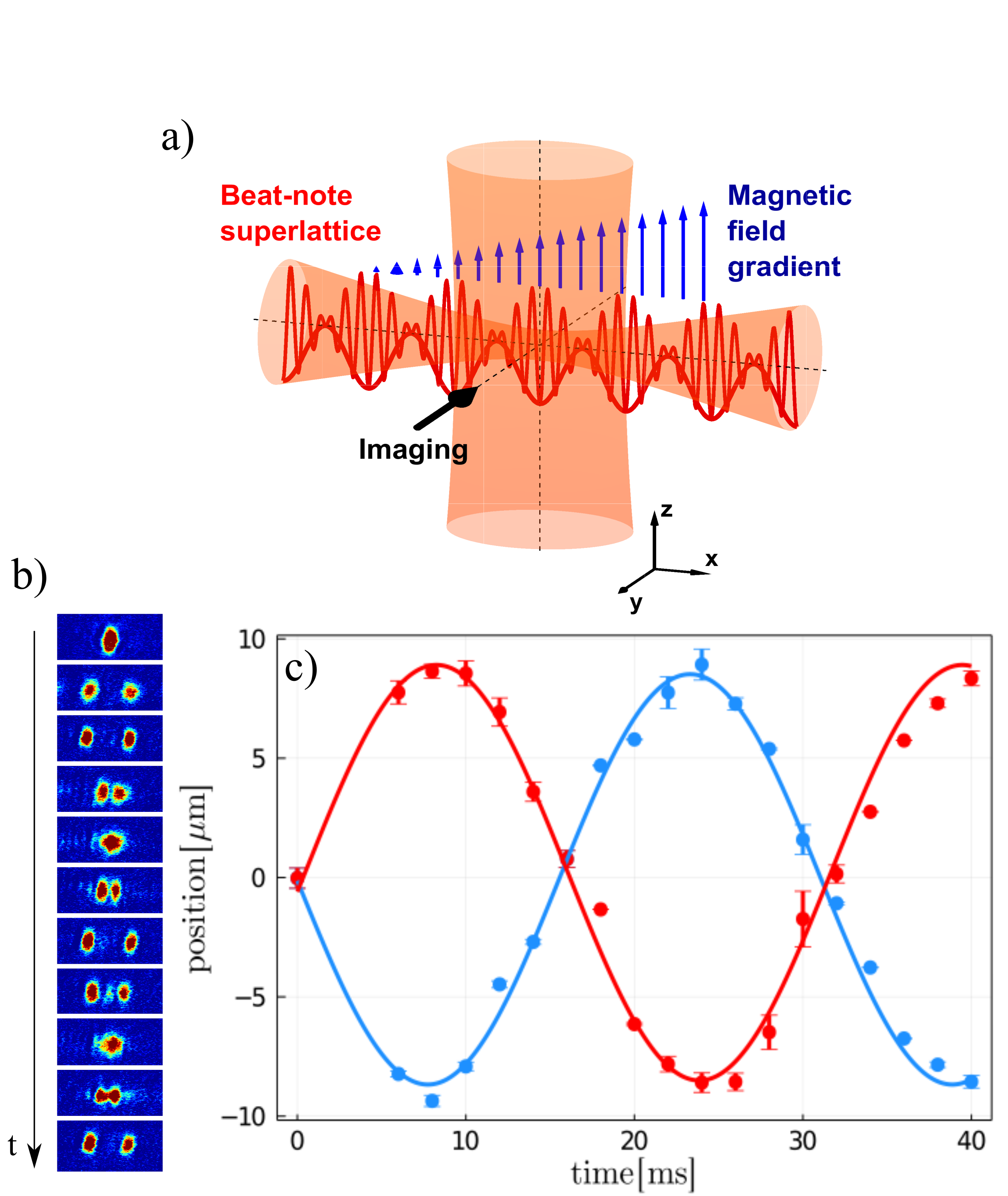}
    \caption{a) Sketch of the optical and magnetic potentials applied on the atoms. b) In-situ absorption images of the diffracted orders $m=\pm1$ during a complete oscillations and corresponding positions as a function of time in c). We observe sinusoidal oscillations with average frequency of $\omega=2\pi \times 31.7(0.8)$ Hz 
    and amplitude of 8.6(0.6) $\mu$m.}
    \label{fig:exp}
\end{center}
\end{figure}

In order to realize an ``ideal" interferometer, we use a Bose-Einstein condensate of $10^4$ $^{39}K$ atoms in the $\ket{f=1, m_f=1}$ state, that features a broad Feshbach resonance around 
400 G \cite{Derrico_Feshbach_2007}. Setting the magnetic field to $B=$350.5(0.5) G effectively cancels the interatomic interactions since the corresponding s-wave scattering length is $|a|<0.05a_0$, $a_0$ being the Bohr radius.\\
We prepare the BEC in a crossed dipole trap created by two red-detuned laser beams, as sketched in Fig. (\ref{fig:exp}a): with a waist of %only 
$17 \mu$m, the horizontal beam provides a tight radial confinement, along $y$ and $z$, of $\omega_r\sim 2\pi\times200$ Hz, while the vertical beam provides the longitudinal harmonic potential, along $x$, with $\omega=2\pi\times(31.7\pm 0.8)$ Hz.   The comparatively large waist of the vertical beam ($100\mu$m) ensures that the deviation from the harmonicity of the potential is below 1$\%$ at a distance of 10 $\mu$m from the center. \\ 
The beat-note superlattice potential along $x$ is generated by overlapping two standing waves $V_1\cos^2 k_1 x + V_2\cos^2 k_2 x$ with wavevectors $k_1=2\pi/1.013 \,\mu\mathrm{m}^{-1}$ and $k_2=2\pi/1.12 \,\mu\mathrm{m}^{-1}$, and $V_1, V_2$ are the lattice depths. 
For $V_1, V_2$ of the order of a recoil energy the BEC experiences an effective lattice potential
$V_{\mathrm{eff}}(1+\sin(k x+\phi))/2$, with
$V_{\mathrm{eff}}=V_1V_2 M/\hbar^2(k_1+k_2)^2$, where $M$ is the atomic mass, the spatial period $2\pi/k= \pi/(k_1-k_2)$ equals $5.3 \,\mu$m, and the phase $\phi$ depends on the relative phase between the two combined lattices \cite{Beatnote_2021}.  Both lasers are frequency locked to the same optical reference cavity with a relative stability of $\sim 10$ kHz, via sideband-locking that allows to tune $\phi$ dynamically
by adjusting the radio-frequency of one sideband. 
Additionally, to investigate the effect of an external force along the $x$ direction we apply a magnetic field gradient, corresponding to an acceleration up to $10^{-3}g$, that displaces the minimum of the resultant harmonic potential in a region of few microns.

\begin{figure}[!htbp]
\begin{center}
    \includegraphics[width=1.0\columnwidth]{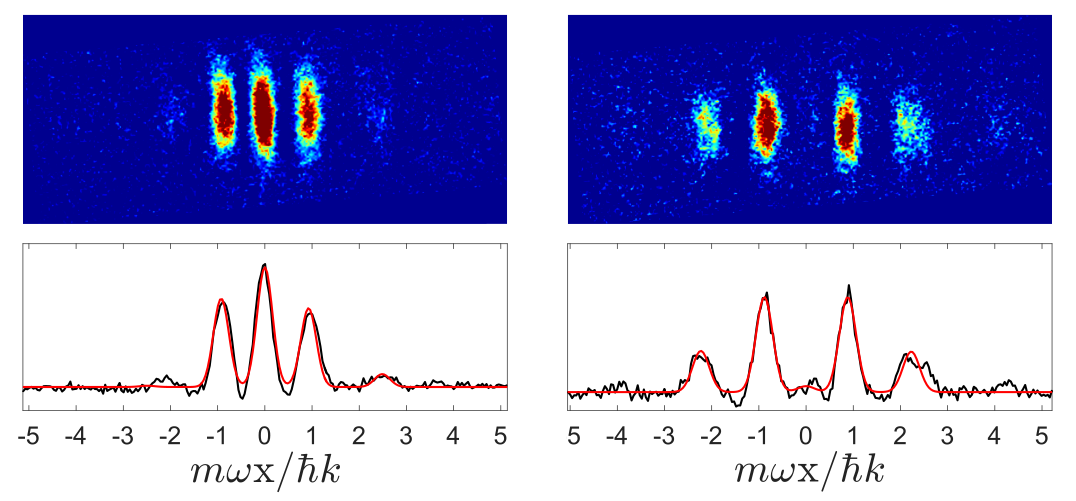} \caption{Absorption image (top) of the momentum distribution and corresponding integrated density profile together with the multi-Gaussian fit (bottom) after the first KD pulse (left) and at the end of the interferometer (right). In both cases     a quarter period of in-trap evolution  occurs between the second KD pulse and imaging. 
    }
    \label{fig:fitted_image}
% fit_blue 
% (182.7677, 182.8836)
% (8.4006, 8.550176)  --> 8.47+-0.07
% (31.4691, 31.7554) --> 31.61+-0.14
% (3.1741, 3.215)
% fit_red
% (182.94989, 183.09311)
% (8.411859, 8.5986870) --> 8.51+-0.09
% (31.682, 32.031)  --> 31.86+-0.17
% (-0.0944, -0.0442)
\end{center}
\end{figure}

In order to measure the longitudinal frequency and confirm that the potential is harmonic, at $t=0$ we shine a pulse of the optical lattice with an effective depth 
$V_{\mathrm{eff}} \simeq 30E_R$ for $120\,\mu$s, where $E_R = \hbar^2 k^2/2M =8.7\,$nK. The pulse is long enough to completely deplete the $m=0$ component. 
We record the images of the two components $m=\pm 1$ via absorption imaging (line of sight along the $y$ direction) and we report their position as a function of time: in Fig. \ref{fig:exp}(b,c), both components display clear sinusoidal oscillations with an amplitude of $8.6(0.6)\mu$m, 
in reasonable agreement with the expected value $\hbar k/M \omega =9.7(0.2)\mu$m.  
Then, we set the KD pulse duration to 80 $\mu$s in order to transfer almost 50$\%$ of the atoms in the $m=\pm1$ orders and we shine the second KD pulse after a half (or full) period to complete the interferometric sequence. We then image the different momentum orders after allowing an additional quarter oscillation in the trap to maximize their spatial separation. We get the atom number in each components, $N_m$, fitting the profiles with a multi-Gaussian function, shown in Fig.\ref{fig:fitted_image}. 

A peculiarity of the beat-note superlattice is represented by the fact that the number of interfering atoms $N$ is coupled to $\beta$. Indeed, beyond the effective potential approximation, the KD pulses diffract atoms also at momentum components associated with the two fundamental optical lattices, i.e. at integer multiples of $2 \hbar k_{1,2}$ \cite{Beatnote_2021} and the atoms of these components are effectively lost for the purpose of the interferometer: due to their large momenta, they are driven in the anharmonic region of the ODT (if not outside). In practice, increasing $\beta$ reduces $N$, the total number of atoms contributing to the interferometer signal; for this reason we work with $\beta<2$.

\begin{figure}[!tbp]
\begin{center}
    \includegraphics[width=1.0\columnwidth]{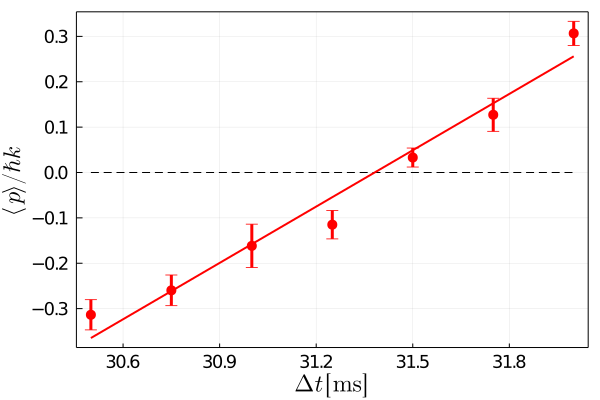}
    \caption{Average momentum at the output of the full-period interferometer as a function of the time separation between the two KD pulses, $\Delta t$. The fit parameter is the harmonic period, identified as the zero-crossing time: $T=(31.38\pm 0.03)$ ms.}
    \label{fig:asymmetry}
\end{center}
\end{figure}

\section{Calibration of the interferometer}

As shown in Fig. \ref{fig:fitted_image} and in agreement with Eq. (\ref{eq:psi_half}), the final momentum distribution is an even function of $m$, independently from the phases $\phi$ and $\delta$, for both the half- and the full-period interferometer; thus, $\langle p \rangle =0$. However, this symmetry breaks if the time separation of the two KD pulses is not exactly half period, i.e. $\Delta t= (1/2+\epsilon)T$. In this case, the $m-th$ momentum component acquires an extra phase, 
$\exp[i \epsilon 2\pi m^2 E_R/\hbar \omega]$, given by the classical action in the time interval between the KD pulses. We exploit this sensitivity to precisely determine the oscillation period, using the full-period interferometer, that is inherently more stable, since it is insensitive to the relative displacement between the trap and the lattice (and thus to external forces). We identify the period as the time separation $\Delta t$ that yields a symmetric momentum distribution; indeed  Fig. \ref{fig:asymmetry} shows that measuring the average momentum $\langle p \rangle$ as a function of $\Delta t$ allows to find the oscillation period with a precision of $10^{-4}$, a factor 27 more accurate than what was possible by measuring the oscillation of the spatial displacement of the wave packets, shown in Fig. \ref{fig:exp}. Then this value is used to set the time separation equal to $T/2$, in the half-period interferometer which is sensitive to external forces.
\begin{figure}[!tbp]
\begin{center}
    \includegraphics[width=1.0\columnwidth]{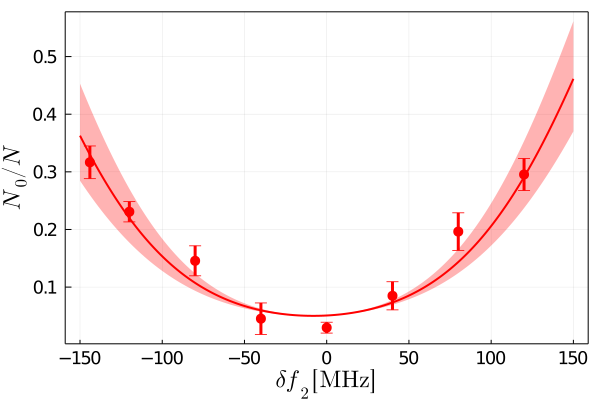}
    \caption{Fraction of atoms in $m=0$ momentum component, $O_1=N_0/N$, at the output of the full-period interferometer as a function of the frequency shift of $k_2$ standing wave during the second KD pulse. The shaded band is  
    the uncertainty of the displacement-vs-current calibration.
    }
    \label{fig:optical_shift}
\end{center}
\end{figure}

\begin{figure}[!tbp]
\begin{center}
    \includegraphics[width=1.0\columnwidth]{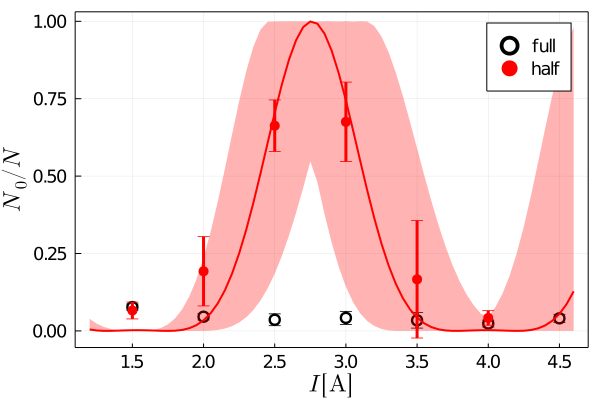}
    \caption{
    Fraction of atoms in $m=0$ momentum component at the interferometer output as a function of the acceleration-induced displacement of the harmonic trap with $\omega=2\pi\times 31.7$ Hz (a), for the half-period (red, solid points)  and full-period (black, open) interferometer. The red line shows the theoretical prediction from Eq. (\ref{eq:psi_half}), with $\beta=1.25$ and an offset phase as a fit parameter, the shaded band is the uncertainty of the displacement-vs-current calibration.
    }
    \label{fig:n0_40_30}
\end{center}
\end{figure}

In addition, with the full-period interferometer we investigate the stability of the relative position of the lattice between the two KD pulses. Eq. (\ref{eq:psi_full}) shows that, at the output of the full-period interferometer, the fraction of 
population at zero momentum  equals 
$J_0^2(2 \beta \cos\delta)$. First, we verify this relation thanks to the dynamical control of the lattice position obtained by frequency shifting one of the two standing waves. We frequency shift the $k_2$ standing wave by $\delta f_2= c\, \delta k_2/(2\pi)$, which displaces the beat-note superlattice by $\delta x_o=L\delta k_2/(k_1-k_2)$, where $L$ is the distance of the atoms from the retro-reflecting mirror. We have calibrated the displacement $\delta x_o$ by {\it in situ} imaging of the position of atoms trapped in the lattice minima, and we obtained $\delta x_o/\delta f_2=(10\pm 1)$ nm/MHz. Thus, during the full-period oscillation we displace the lattice potential by applying the frequency shift $\delta f_2$, which modifies $\phi_2$ by $k \delta x_o$, and we measure $O_1$ as a function of $\delta f_2$: as shown in Fig. \ref{fig:optical_shift} data are well in agreement with the predicted behaviour. 
Since repeated measurements with the full-period, unshifted, interferometer show that the $O_1$ is constant within $0.017$ (standard deviation, see Fig.\ref{fig:n0_40_30}),  we conclude that the relative displacement between the two KD pulses, over the time scale of one oscillation period, i.e. approximately 30 ms, is bounded to be below $0.2\,\mu$m.

\section{Measurement of the applied force}

With these results in hand, we proceed to measure a real force/acceleration. Since the \ktn\ atoms feature a magnetic moment approximately equal to $0.5 \mu_B$ ($\mu_B$ Bohr magneton) around 350 G, we impart a force along the direction of the lattice by applying an external magnetic field gradient produced  with a pair of coils in anti-Helmholtz configuration. 
The resulting uniform force induces a displacement of the harmonic trap  proportional to the coils current,
which affects the momentum populations at the interferometer output: the displacement-versus-current conversion has been separately calibrated to be $\eta=(1.08\pm0.13)\mu$m/A.

In Fig. \ref{fig:n0_40_30}(a) we report the measured $O_1$ observable as a function of the coil current, with error bars corresponding to the statistical standard deviation for typically 5 repetitions of each data point. The error bar of $\Delta O_1=0.1$, at the maximum slope of the fit curve, yields a resolution of $\Delta I=0.05\,$ A, from which we obtain $\Delta a = 1.8(0.2)\cdot 10^{-4} g$, where the uncertainty is actually dominated by the systematic error due to calibration factor $\eta$.

\begin{figure}[!tbp]
\begin{center}
    \includegraphics[width=1.0\columnwidth]{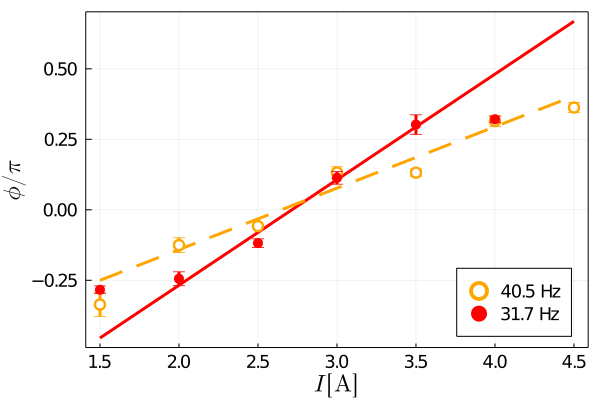}
    \caption{Interferometric phase $\phi$ derived from fitting the momentum fractional populations of momentum components with $J_m^2(2\beta \sin(\phi))$, for harmonic frequencies equal to $31.7$ Hz (red, solid points) and
    $40.5$ Hz (orange, open); lines are linear fits with slopes $(1.18 \pm 0.09)$ rad/A (red, solid, excluding extreme points) and $(0.68 \pm 0.08)$ rad/A (orange, dash).}
    \label{fig:fitted_phase}
\end{center}
\end{figure}

We also assess the acceleration resolution in a complementary manner, i.e. by extracting the interferometric phase $\phi$ from a fit of the populations of all momentum components, for each value of the applied force. 
This phase is expected to depend linearly on the force, hence on the current: indeed this is what we observe in Fig. \ref{fig:fitted_phase}. From the average error bar of these data, $\langle \Delta \phi\rangle=0.06$, we obtain the acceleration sensitivity $\Delta a =(\omega^2/k) \langle \Delta \phi \rangle  = 2.2\cdot 10^{-4} g$, consistent with the one above.   

These values must be compared to the Cramer-Rao bound reported in \cite{Li_Multimode_2014}, i.e. $\Delta a_{CR} =  (\omega^2/\beta k) 1/(\sqrt{8pN})$ where $N$ is the number of atoms, and $p$ is the number of repeated measurements.
In our experiment, $N\simeq 5\cdot 10^3$ and $p$ is typically 4, thus $1/\sqrt{8pN} = 2.5\cdot 10^{-3}$, which is approximately a factor 40 smaller than our experimental $\Delta O_1$. In other words, our resolution does not reach the standard quantum limit corresponding to $\Delta a_{CR} = 4\cdot 10^{-6} g$. While a 
more systematic investigation is needed, we believe that the main cause for the sub-optimal performance is the pointing instability of the ODT beams. Indeed, any technical displacement of the trap minimum is indistinguishable from those induced by external accelerations; the interferometric sequence lasts only 16 ms but the sample preparation (dead time) takes approximately half a minute, and slow drifts of the ODT beams occur over this time scale. The measured sensitivity is equivalent to a displacement of $0.2\,\mu$m, reasonably of the same order of the slow drifts of the ODT position. 

The above discussion shows that lowering the harmonic frequency improves the resolution. However, in our setup we can only marginally increase -- and not decrease -- the trapping frequency, to keep under control the anharmonicity of the potential. Thus, we repeated the measurement with $\omega = 2\pi \times (40.5\pm 0.5)$ Hz and, as expected, 
the measured interferometric phase is less sensitive to the applied force (see Fig. \ref{fig:fitted_phase}).

\section{Conclusions}
In conclusion, we have performed a proof-of-principle demonstration of a multimode interferometer in a harmonic trap based on KD diffraction pulses. We have shown that external accelerations are detected from the displacement induced on the harmonic trap with respect to the KD lattice.   
With a relatively low number of atoms $N\simeq 5\cdot 10^3$ and a harmonic frequency of approximately 32 Hz, we showed a sensitivity $\delta a \simeq 2\cdot 10^{-4}g$. Our result is a factor 40 away from the Cramer-Rao bound due to instabilities of the position of the harmonic potential with respect to the lattice. 

For a better insight on the potential performance of the KD interferomter, we rewrite Eq.(\ref{eq:min_delta_a}) as 
\begin{equation}
    \frac{\delta a}{a} \simeq \frac{\hbar \omega}{M a A} \Delta O_1,
\end{equation}
to show that the relative sensitivity is inversely proportional to the potential energy difference of the external force at distances equal to the oscillation amplitude of the atoms with the largest momentum, $A= \beta \hbar k/(M\omega)$. This expression shows that 
ours represents a simple method to enhance the sensitivity of a trapped atom interferometer by enlarging the spatial separation between the modes involved.  In particular this is done by: {\it (i)} reducing the harmonic trapping frequency, {\it (ii)} increasing the number of momentum components with KD pulses of enhanced intensity and/or duration \footnote{In an ODT the oscillations of the largest momentum components are harmonic only if their amplitude is much smaller than the beams waist $w$, i.e. $\beta \hbar k/(M \omega) \ll w$, which constraints the achievable sensitivity.}.

In the future it will be interesting to explore the performance of the sensor using a harmonic magnetic confinement along the direction of the lattice, that is more stable in position than the ODT and features a harmonic region much larger. For example, using $10^5$ atoms in a magnetic trap with frequency $\approx 1$ Hz and with oscillation amplitudes of $\approx 1$ mm, we expect an improvement of the sensitivity by several orders of magnitude, up to $\sim 10^{-8}g$. In a magnetic trap, the inhomogeneous magnetic field experienced by the atoms along the oscillation is a concern since it changes the interatomic interaction strength: for $^{39}$K atoms, the above combination of magnetic confinement and oscillation amplitudes implies that the variation of magnetic field is $\approx 10$ mG, corresponding to a negligible variation of the scattering length $\approx 0.006 a_0$.  
Interestingly, the KD interferometer could be also implemented with spin-polarized fermionic atoms that, at low temperature, are naturally non-interacting, provided that the KD lattice spacing is chosen smaller than the coherence length of the atomic sample \cite{roati2004}. 

\acknowledgements
We acknowledge fruitful discussions with A. Smerzi and we thank  M. Prevedelli for the critical reading of the manuscript.
This work was supported by the projects TAIOL of QuantERA ERA-NET Cofund in Quantum Technologies (Grant Agreement n. 731473) and QOMBS of FET Flagship on Quantum Technologies (Grant Agreement n. 820419),
implemented within the European Union’s Horizon 2020 Programme.

\bibliography{KDinter}% Produces the bibliography via BibTeX.

\end{document}